\documentclass[twocolumn,showpacs,preprintnumbers,amsmath,amssymb]{revtex4}

\usepackage{graphicx}
\usepackage{dcolumn}
\usepackage{bm}

\begin{document}


\title{\large Probing a Kondo correlated quantum dot with spin spectroscopy}

\author{D. Kupidura$^1$}
\author{M.~C. Rogge$^1$}
\email{rogge@nano.uni-hannover.de}
\author{M. Reinwald$^2$}
\author{W. Wegscheider$^2$}
\author{R.~J. Haug$^1$}
 \affiliation{$^1$Institut f\"ur Festk\"orperphysik, Universit\"at Hannover,
  Appelstra\ss{}e 2, D-30167 Hannover, Germany \\ $^2$Angewandte und Experimentelle Physik, Universit\"at
Regensburg, D-93040 Regensburg, Germany}

\date{\today}

\begin{abstract}
We investigate Kondo effect and spin blockade observed on a
many-electron quantum dot and study the magnetic field dependence.
At lower fields a pronounced Kondo effect is found which is
replaced by spin blockade at higher fields. In an intermediate
regime both effects are visible. We make use of this combined
effect to gain information about the internal spin configuration
of our quantum dot. We find that the data cannot be explained
assuming regular filling of electronic orbitals. Instead spin
polarized filling seems to be probable.
\end{abstract}

\maketitle



Quantum dot systems containing confined electrons have been
studied intensively both theoretically \cite{Beenakker-91} and
experimentally \cite{Kouwenhoven-01} during the last decades. They
have proven to be excellent systems to explore the physics of
confined charge carriers. Beyond Coulomb blockade
\cite{Beenakker-91,Kastner-93} many effects have been found
revealing quantum mechanical and spin properties of confined
electrons, among them Kondo effect and spin blockade. The Kondo
effect is caused by correlated cotunnelling events
\cite{Sasaki-00,Goldhaber-Gordon-98} resulting in enhanced
conductance when sequential tunnelling is Coulomb blocked. The
underlying mechanism is due to the formation of a spin singlet
between conduction electrons and the localized spin in the quantum
dot. Spin blockade as introduced by Ciorga et al. \cite{Ciorga-00}
appears in lateral dot systems when applying a magnetic field
perpendicular to the sample surface. Spin dependent transport
suppression is found as a consequence of electron injection from
spin polarized edge states.





In this work we investigate the interplay between these two
effects, as they appear in the same sample in slightly different
magnetic field regions. We study the magnetoconductance for Kondo
effect and spin blockade individually and the transition between
them. Results from both effects are then combined to gain
information about the internal spin configuration of the quantum
dot.

\begin{figure}
\centering
\includegraphics[scale=1]{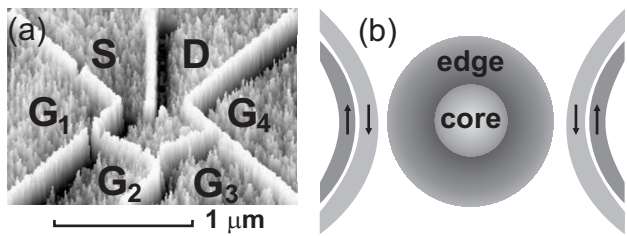}
\caption{a) AFM picture of our device with Source (S) and Drain
(D) leads and gates (G$_1$ to G$_4$). b) Schematic picture: under
the influence of a magnetic field the dot can be represented as
edge and core areas for the two lowest Landau levels. The dot is
coupled to Source and Drain represented here by spin polarized
edge states.} \label{fig1}
\end{figure}

Our device is defined by terms of Local Anodic Oxidation (LAO)
\cite{Ishii-95} on a GaAs/AlGaAs heterostructure containing a
2-dimensional electron system (2DES) $37$~nm below the surface.
The 2DES has an electron density of $n_{e}=
3.94\cdot10^{15}$~m$^{-2}$ and mobility of $\mu_{e}=
52.7$~m$^{2}$V$^{-1}$s$^{-1}$ at $4.2$~K. Figure \ref{fig1}a)
shows an Atomic Force Microscope (AFM) picture of our dot with
leads Source (S) and Drain (D) being the electron reservoirs and
in plane gates (G$_1$ - G$_4$) respectively. Gates G$_1$ and G$_4$
are designed to control the coupling to the leads while gates
G$_2$ and G$_3$ control the electrostatic potential of the dot.
The measurements are performed in a He$^3$/He$^4$ dilution
refrigerator at $50$~mK with a magnetic field $B$ applied
perpendicular to the 2DES. Standard lock in technique is used to
record the differential conductance $G$ as a function of gate
voltages and magnetic field.

Figure \ref{fig1}b) shows a simplified model of our system in the
relevant field regime. The properties of the dot are basically
characterized by the two lowest Landau levels formed at edge and
core of the dot (filling factor $\nu_{dot} >2$). The 2-dimensional
leads are subject to the Quantum Hall Effect and thus develop spin
polarized edge channels that get separated in space with
increasing $B$. The visibility of Kondo effect and spin blockade
depends on this separation and thus on the strength of the
magnetic field.

\begin{figure}
\centering
\includegraphics[scale=1]{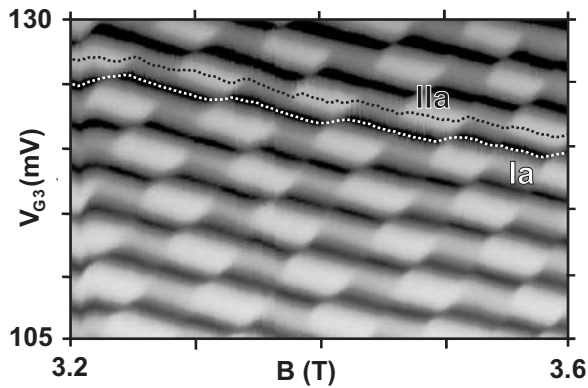}
\caption{Differential conductance $G$ as a function of magnetic
field $B$ and gate voltage $V_{G3}$. Dark color stays for high and
white for low conductance. A chessboard pattern is observed caused
by Kondo conductance in the Coulomb valleys between individual
Coulomb peaks. The Kondo effect is modulated by electron number
and magnetic field.} \label{fig2}
\end{figure}

For relatively low fields up to approx. $3.5$~T the leads can
still be considered as unpolarized. Here we find Kondo physics.
The Kondo effect is induced by a common state of electrons on the
dot and electrons in the leads. If - high tunnel coupling presumed
- the total spin of the dot is $S=1/2$ a singlet state can be
formed with the leads allowing cotunnelling events even if
sequential tunnelling is Coulomb blocked. As a consequence finite
transport is observed between individual Coulomb peaks. Since
transport via this Kondo state involves spin flips both spins must
be present in the leads.

Figure \ref{fig2} shows a measurement in the relevant Kondo
regime. The differential conductance $G$ is measured as a function
of the voltage applied to gate 3 and the magnetic field (dark for
high $G$, white for low). Several Coulomb peaks are visible. Their
peak positions e.g. the ground state energies exhibit a zigzag
behavior as the field increases. Negative gradients correspond to
transport through the edge of the dot while transport through the
center leads to a positive gradient. In between we find a regular
pattern of enhanced conductance caused by a Kondo effect involving
the edge of the dot. The center of the dot is not involved since
the tunnel coupling to the leads is not high enough. Thus
transport via a Kondo singlet state corresponds to a total spin
$S=1/2$ at the edge of the dot. This spin depends on the number of
electrons at the edge which can be changed in two different ways.
An electron can enter the edge either coming from the leads or
coming from the dot center. The first mechanism is induced by
increasing the gate voltage increasing the total number of
electrons. Therefore a regular modulation of the Kondo effect is
found as a function of $V_{G3}$. The second mechanism is used when
increasing the magnetic field. With increasing $B$ electrons are
redistributed from the second Landau level in the center to the
first Landau level at the edge. This leads to a Kondo modulation
as a function of $B$ and overall to a chessboard pattern as
reported in Refs. \cite{Fühner-02,Keller-01,Stopa-03}.

\begin{figure}
\centering
\includegraphics[scale=1]{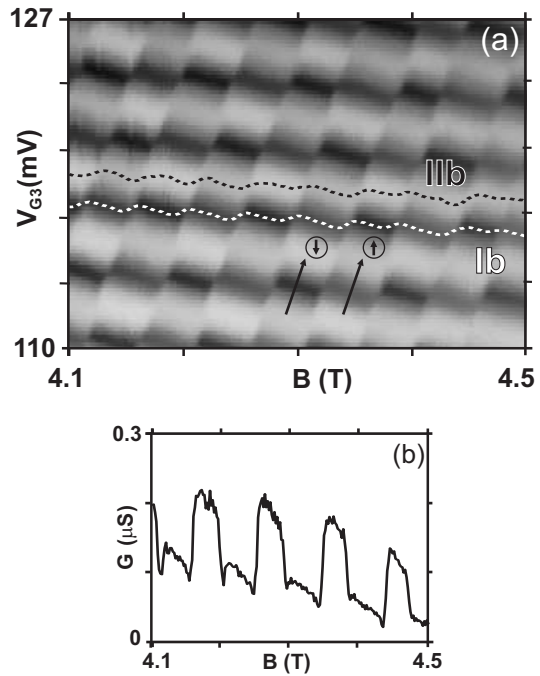}
\caption{ (a) At higher fields spin polarized edge states in the
leads give rise to the spin blockade. Transport with spin up
electrons is suppressed compared to spin down transport. The
arrows denote charging with the assigned spin detectable with the
amplitude of the corresponding Coulomb peak. (b) Peak amplitude
for peak Ib. The bimodal behavior due to spin blockade is clearly
visible.} \label{fig3}
\end{figure}

Above $3.5$~T the spatial separation of the edge channels gets
prominent and the leads are more and more spin polarized leading
to a pronounced spin blockade. This is visible in Fig.
\ref{fig3}(a). Again several Coulomb peaks are visible as a
function of $V_{G3}$ and $B$ but this time for a higher field
range. The zigzag pattern of the peak position still reflects the
energy of the ground states located at the center of the dot
(positive slope) or at the edge (negative slope). In addition a
bimodal behavior of the peak amplitude is found when transport
occurs through the edge. This is shown in Fig. \ref{fig3}(b). The
peak amplitude for the peak marked as Ib in (a) is plotted as a
function of $B$. Since the edge states in the leads are spatially
separated the overlap of the wavefunctions between lead and dot
for the energetically higher spin state (spin up in the following)
is reduced and transport is blocked. In contrast electrons with
the lower spin state (spin down) cause an unsuppressed transport.
This effect allows to detect the spin of an electron added to the
dot. Crossing a Coulomb peak in gate voltage direction with an
unsuppressed amplitude leads to the addition of an electron with
spin down while the addition via a low amplitude corresponds to a
spin up electron (see arrows in Fig. \ref{fig3}(a)).

\begin{figure}
\centering
\includegraphics{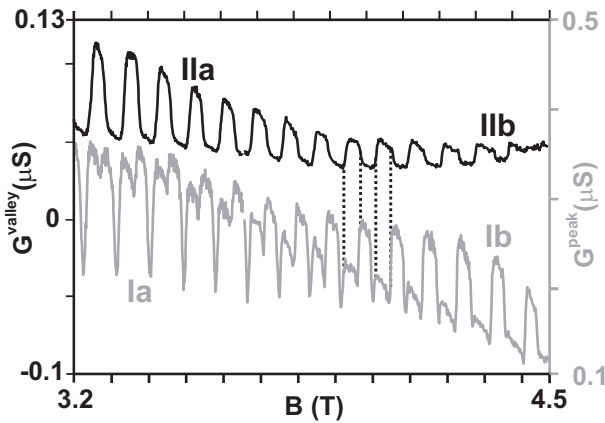}
\caption{Peak amplitude along Coulomb peak Ia/Ib (gray curve) and
conductance in the Coulomb valley IIa/IIb above (black curve).
While the bimodal behavior along the peak increases with magnetic
field due to spin blockade the oscillations due to Kondo effect in
the valley vanish. Around $4$~T both effects are visible and
comparable. A high amplitude at the peak goes along with a
suppressed conductance in the valley (see dotted lines).}
\label{fig4}
\end{figure}

So far both regimes were investigated separately. In the following
we will concentrate on the interplay between Kondo effect and spin
blockade. The peak amplitudes of the peaks marked as Ia and Ib in
Figs. \ref{fig2} and \ref{fig3} are plotted in gray as a function
of magnetic field in Fig. \ref{fig4}. The strength of the bimodal
behavior reflects the strength of the spin blockade. At $3.2$~T
almost no oscillation is visible. The leads are still unpolarized
and thus no spin blockade is observed. With increasing $B$ the
edge states start to split and the spin blockade gets visible.
Above $4$~T it reaches its maximum. The black curve represents the
conductance in the Coulomb valley (IIa and IIb in Fig. \ref{fig2}
and \ref{fig3}) above the depicted peak. At $3.2$~T a strong
oscillation is observed due to a pronounced Kondo effect.
Increasing the field separates the edge states. This leads to a
suppression of the Kondo effect, since the transport over the
Kondo state involves spin flips requiring both spin orientations
in the leads. At $4.5$~T the Kondo effect is strongly suppressed.

\begin{figure}
\centering
\includegraphics{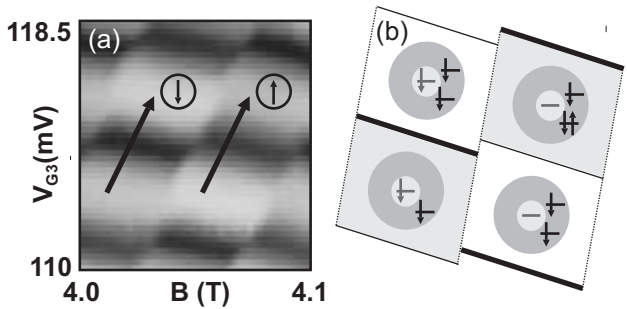}
\caption{(a) increased region of Fig. \ref{fig3}(a) showing both
Kondo effect and spin blockade. (b) model of the spin
configuration of the dot gained from the combination of results
from Kondo effect and spin blockade. The coincidence of low
conductance along a peak and high conductance in the valley above
is explained with spin polarized filling of electrons.}
\label{fig5}
\end{figure}

In an intermediate regime around $4$~T both effects are visible
simultaneously. Thus the results from both effects can be
combined. We find that a high peak amplitude corresponding to spin
down transport is accompanied by a suppressed Kondo effect in the
Coulomb valley above. Therefore the addition of an electron to the
energetically lower spin down state does not lead to a total spin
$S=1/2$ required for Kondo conductance. Instead the Kondo effect
reflecting a total spin of $S=1/2$ appears above a low peak
amplitude when an electron with the energetically higher spin up
state is added. This discrepancy shows that the electronic state
structure within our dot cannot be described in terms of
noninteracting electrons. A simple model with regular filling of
dot orbitals featuring a spin singlet phase is not applicable.
Instead the addition of electrons is realized favoring spin
polarized configurations, i.e. a sort of Hund's rule.

We propose an explanation based on theoretical results inspired by
an experiment given by Ciorga et al. \cite{Ciorga-02}. They
investigated the $\nu_{dot}=2$ transition in a lateral quantum dot
featuring spin blockade as well but no Kondo effect. A phase
transition from a spin singlet phase towards a spin polarized
phase was found depending on the number of electrons. For electron
numbers below approx. $30$ the spin singlet phase was observed
with a regular filling of dot orbitals. Above this critical number
spin polarized configurations were observed.

We find that the model developed by Ciorga et al. for the internal
spin configuration above the critical electron number can be
applied to our results which fits the assumption of having a high
electron number $N\approx 160$. The model is shown in Fig.
\ref{fig5}. Four "Kondo fields" are shown with Kondo conductance
in the lower left and the upper right ((a) shows the measurement
and (b) the model). Thus in these fields we assume the total spin
of the edge to be $S=1/2$. This is explained with the highest
orbital in the edge to be half filled with a spin down electron.
Going from the lower left to the upper left a Coulomb peak is
crossed with a high amplitude. Thus the half filled orbital is not
filled with a spin up electron. Instead the next orbital is
entered with a spin down electron. The total spin of the edge is
$S=1$ and the Kondo effect disappears. If now the magnetic field
is increased, an electron is redistributed from the center to the
edge and the first half filled orbital at the edge is filled
leaving one half filled orbital with the total spin $S=1/2$. The
Kondo effect is restored.

We want to mention that the results obtained here may have a
universal character. Very similar results were obtained by Keller
et al. \cite{Keller-01}. A Kondo induced chessboard pattern was
observed. The peak amplitude which behaves the same way as in our
sample was plotted but no attention was paid to this effect. The
chessboard was explained with regular filling featuring a spin
singlet phase. This is in our opinion not valid. Instead an
electron number of approx. $50$ supports our assumption of spin
polarized filling. Measurements done by F\"uhner et al.
\cite{Fühner-02} for an electron number of approx. $180$ showing
the magnetic flux dependence of the chessboard pattern also
confirm our results. Other chessboard measurements combined with
calculations were performed by Stopa et al. \cite{Stopa-03}. With
an electron number below $30$ they assumed a regular filling.
Their calculations might therefore not be applicable for
chessboard patterns at higher electron numbers. Regular filling
was also obtained in \cite{Rogge-05}. The electron number is
slightly above $50$ but seems still to be low enough for the spin
singlet phase probably due to different dot parameters.

In conclusion we investigate Kondo effect and spin blockade in a
lateral quantum dot depending on the strength of a magnetic field
applied perpendicular to the sample surface. At lower fields we
observe a Kondo induced chessboard pattern modulated by the number
of electrons and the magnetic field. At higher fields the leads
get spin polarized and spin blockade sets in making the electronic
spin detectable. The Kondo conductance vanishes here since both
spin orientations in the leads are needed for the Kondo effect. In
intermediate fields both effects are observed. A combined analysis
reveals the internal spin configuration of the dot. We find that
the measurements cannot be described by a regular filling of
electronic orbitals within a model of noninteracting electrons.
Instead for this high electron number a different approach
featuring spin polarized filling with a sort of Hund's rule must
be assumed. The comparison with other results challenges former
interpretations of chessboard patterns and yields a consistent
picture with the dot phase depending on the electron number.

This work has been supported by BMBF.




\begin{thebibliography}{20}
\expandafter\ifx\csname
natexlab\endcsname\relax\def\natexlab#1{#1}\fi
\expandafter\ifx\csname bibnamefont\endcsname\relax
  \def\bibnamefont#1{#1}\fi
\expandafter\ifx\csname bibfnamefont\endcsname\relax
  \def\bibfnamefont#1{#1}\fi
\expandafter\ifx\csname citenamefont\endcsname\relax
  \def\citenamefont#1{#1}\fi
\expandafter\ifx\csname url\endcsname\relax
  \def\url#1{\texttt{#1}}\fi
\expandafter\ifx\csname
urlprefix\endcsname\relax\def\urlprefix{URL }\fi
\providecommand{\bibinfo}[2]{#2}
\providecommand{\eprint}[2][]{\url{#2}}

\bibitem[{\citenamefont{Beenakker}(1991)\citenamefont{Beenakker}}]{Beenakker-91}
\bibinfo{author}{\bibfnamefont{C.~W.~J.} \bibnamefont{Beenakker}},
  \bibinfo{journal}{Phys. Rev. B} \textbf{\bibinfo{volume}{44}},
  \bibinfo{pages}{1646} (\bibinfo{year}{1991}).


\bibitem[{\citenamefont{Kouwenhoven et~al.}(2001)\citenamefont{Kouwenhoven, Austing, and Tarucha}}]{Kouwenhoven-01}
\bibinfo{author}{\bibfnamefont{L.~P.} \bibnamefont{Kouwenhoven}},
\bibinfo{author}{\bibfnamefont{D.~G.}~\bibnamefont{Austing}},
\bibnamefont{and} \bibinfo{author}{\bibfnamefont{S.}~\bibnamefont{Tarucha}},
  \bibinfo{journal}{Rep. Prog. Phys.} \textbf{\bibinfo{volume}{64}},
  \bibinfo{pages}{701} (\bibinfo{year}{2001}).

\bibitem[{\citenamefont{Kastner}(1993)\citenamefont{Kastner}}]{Kastner-93}
\bibinfo{author}{\bibfnamefont{M.~A.} \bibnamefont{Kastner}},
  \bibinfo{journal}{Physics Today},
  ~\bibinfo{pages}{24} (\bibinfo{year}{Jan 1993}).



\bibitem[{\citenamefont{Sasaki et~al.}(2000)\citenamefont{Sasaki, Franceschi, Elzerman, van der Wiel, Eto, Tarucha, and Kouwenhoven}}]{Sasaki-00}
\bibinfo{author}{\bibfnamefont{S.} \bibnamefont{Sasaki}},
\bibinfo{author}{\bibfnamefont{S.}~\bibnamefont{De Franceschi}},
\bibinfo{author}{\bibfnamefont{J.~M.}~\bibnamefont{Elzerman}},
\bibinfo{author}{\bibfnamefont{W.~G.}~\bibnamefont{van der Wiel}},
\bibinfo{author}{\bibfnamefont{M.}~\bibnamefont{Eto}},
\bibinfo{author}{\bibfnamefont{S.}~\bibnamefont{Tarucha}},
\bibnamefont{and} \bibinfo{author}{\bibfnamefont{L.~P.} \bibnamefont{Kouwenhoven}},
  \bibinfo{journal}{Nature.} \textbf{\bibinfo{volume}{405}},
  \bibinfo{pages}{764} (\bibinfo{year}{2000}).

\bibitem[{\citenamefont{Goldhaber-Gordon et~al.}(1998)\citenamefont{Goldhaber-Gordon, Shtrikman, Mahalu, Abusch-Magder, Meirav, and Kastner}}]{Goldhaber-Gordon-98}
\bibinfo{author}{\bibfnamefont{D.} \bibnamefont{Goldhaber-Gordon}},
\bibinfo{author}{\bibfnamefont{Hadas}~\bibnamefont{Shtrikman}},
\bibinfo{author}{\bibfnamefont{D.}~\bibnamefont{Mahalu}},
\bibinfo{author}{\bibfnamefont{David}~\bibnamefont{Abusch-Magder}},
\bibinfo{author}{\bibfnamefont{U.}~\bibnamefont{Meirav}},
\bibnamefont{and} \bibinfo{author}{\bibfnamefont{M.~A.} \bibnamefont{Kastner}},
  \bibinfo{journal}{Nature} \textbf{\bibinfo{volume}{391}},
  \bibinfo{pages}{156} (\bibinfo{year}{1998}).



\bibitem[{\citenamefont{Ciorga et~al.}(2000)\citenamefont{Ciorga, Sachrajda, Hawrylak, Gould, Zawadzki, Jullian, Feng, and Wasilewski}}]{Ciorga-00}
\bibinfo{author}{\bibfnamefont{M.} \bibnamefont{Ciorga}},
\bibinfo{author}{\bibfnamefont{A.~S.} \bibnamefont{Sachrajda}},
\bibinfo{author}{\bibfnamefont{P.} \bibnamefont{Hawrylak}},
\bibinfo{author}{\bibfnamefont{C.} \bibnamefont{Gould}},
\bibinfo{author}{\bibfnamefont{P.} \bibnamefont{Zawadzki}},
\bibinfo{author}{\bibfnamefont{S.} \bibnamefont{Jullian}},
\bibinfo{author}{\bibfnamefont{Y.} \bibnamefont{Feng}},
\bibnamefont{and} \bibinfo{author}{\bibfnamefont{Z.} \bibnamefont{Wasilewski}},
  \bibinfo{journal}{Phys. Rev. B} \textbf{\bibinfo{volume}{61}},
  \bibinfo{pages}{R16315} (\bibinfo{year}{2000}).


\bibitem[{\citenamefont{Ishii et~al.}(1995)\citenamefont{Ishii, and Matsumoto}}]{Ishii-95}
\bibinfo{author}{\bibfnamefont{M.} \bibnamefont{Ishii}},
\bibnamefont{and} \bibinfo{author}{\bibfnamefont{K.} \bibnamefont{Matsumoto}},
  \bibinfo{journal}{JJAP} \textbf{\bibinfo{volume}{34}},
  \bibinfo{pages}{1329} (\bibinfo{year}{1995}).



\bibitem[{\citenamefont{F\"uhner et~al.}(2002)\citenamefont{F\"uhner, Keyser, and Haug}}]{Fühner-02}
\bibinfo{author}{\bibfnamefont{C.} \bibnamefont{F\"uhner}},
\bibinfo{author}{\bibfnamefont{U.~F.} \bibnamefont{Keyser}},
\bibinfo{author}{\bibfnamefont{R.~J.} \bibnamefont{Haug}},
\bibinfo{author}{\bibfnamefont{D.} \bibnamefont{Reuter}},
\bibnamefont{and} \bibinfo{author}{\bibfnamefont{A.~D.} \bibnamefont{Wieck}},
  \bibinfo{journal}{Phys. Rev. B} \textbf{\bibinfo{volume}{66}},
  \bibinfo{pages}{161305(R)} (\bibinfo{year}{2002}).

\bibitem[{\citenamefont{Keller et~al.}(2001)\citenamefont{Keller, Wilhelm, Schmid, Weis, von Klitzing, and Eberl}}]{Keller-01}
\bibinfo{author}{\bibfnamefont{M.} \bibnamefont{Keller}},
\bibinfo{author}{\bibfnamefont{U.} \bibnamefont{Wilhelm}},
\bibinfo{author}{\bibfnamefont{J.} \bibnamefont{Schmid}},
\bibinfo{author}{\bibfnamefont{J.} \bibnamefont{Weis}},
\bibinfo{author}{\bibfnamefont{K.} \bibnamefont{von Klitzing}},
\bibnamefont{and} \bibinfo{author}{\bibfnamefont{K.} \bibnamefont{Eberl}},
  \bibinfo{journal}{Phys. Rev. B} \textbf{\bibinfo{volume}{64}},
  \bibinfo{pages}{033302} (\bibinfo{year}{2001}).


\bibitem[{\citenamefont{Stopa et~al.}(2003)\citenamefont{Stopa, van der Wiel, De Franceschi, Tarucha, and Kouwenhoven}}]{Stopa-03}
\bibinfo{author}{\bibfnamefont{M.} \bibnamefont{Stopa}},
\bibinfo{author}{\bibfnamefont{W.~G.} \bibnamefont{van der Wiel}},
\bibinfo{author}{\bibfnamefont{S.} \bibnamefont{De Franceschi}},
\bibinfo{author}{\bibfnamefont{S.} \bibnamefont{Tarucha}},
\bibnamefont{and} \bibinfo{author}{\bibfnamefont{L.~P.} \bibnamefont{Kouwenhoven}},
  \bibinfo{journal}{Phys. Rev. Lett.} \textbf{\bibinfo{volume}{91}},
  \bibinfo{pages}{046601} (\bibinfo{year}{2003}).



\bibitem[{\citenamefont{Ciorga et~al.}(2002)\citenamefont{Ciorga, Wensauer, Pioro-Ladriere, Korkusinski, Kyriakidis, Sachrajda, and Hawrylak}}]{Ciorga-02}
\bibinfo{author}{\bibfnamefont{M.} \bibnamefont{Ciorga}},
\bibinfo{author}{\bibfnamefont{A.} \bibnamefont{Wensauer}},
\bibinfo{author}{\bibfnamefont{M.} \bibnamefont{Pioro-Ladriere}},
\bibinfo{author}{\bibfnamefont{M.} \bibnamefont{Korkusinski}},
\bibinfo{author}{\bibfnamefont{J.} \bibnamefont{Kyriakidis}},
\bibinfo{author}{\bibfnamefont{A.~S.} \bibnamefont{Sachrajda}},
\bibnamefont{and} \bibinfo{author}{\bibfnamefont{P.} \bibnamefont{Hawrylak}},
  \bibinfo{journal}{Phys. Rev. Lett.} \textbf{\bibinfo{volume}{88}},
  \bibinfo{pages}{256804} (\bibinfo{year}{2002}).


\bibitem[{\citenamefont{Rogge et~al.}(2005)\citenamefont{Rogge, Cavaliere, Sassetti, Haug, and Kramer}}]{Rogge-05}
\bibinfo{author}{\bibfnamefont{M.~C.} \bibnamefont{Rogge}},
\bibinfo{author}{\bibfnamefont{F.} \bibnamefont{Cavaliere}},
\bibinfo{author}{\bibfnamefont{M.} \bibnamefont{Sassetti}},
\bibinfo{author}{\bibfnamefont{R.~J.} \bibnamefont{Haug}},
\bibnamefont{and} \bibinfo{author}{\bibfnamefont{B.} \bibnamefont{Kramer}},
  \bibinfo{journal}{cond-mat/0507036}.




\end{thebibliography}



\end{document}